\documentstyle[12pt,a4]{article}
\newcommand{\Bomega}{\mbox{\boldmath $\omega$\unboldmath}}
\newcommand{\Bpsi}{\mbox{\boldmath $\psi$\unboldmath}}
\newcommand{\diag}{\,\mbox{\rm diag}\,}
\newcommand{\const}{\,\mbox{\rm const}\,}

\begin{document}

\title{Nontrivial Fermion States in Supersymmetric
Minisuperspace\thanks{Invited lecture presented at the `First Mexican School
on Gravitation and Mathematical Physics',\newline Guanajuato, GTO., Mexico,
Dec. 12--16, 1994}}
\author{Andr\'as Csord\'as\thanks{Permanent address: Research Institute
for Solid State Physics, P.O. Box 49, H1525 Budapest, Hungary} and
Robert Graham}
\date{Fachbereich Physik, Universit\"at-Gesamthochschule Essen\\
45117 Essen\\ Germany}
\maketitle

\begin{abstract}
In this lecture supersymmetric minisuperspace models of any Bianchi type
within class A of the classification of Ellis and McCallum are considered.
The algebra
of the supersymmetry generators, the Lorentz generators, the diffeomorphism
generators and the Hamiltonian generator is determined explicitely and found
to close.
Different from earlier work it is established that physical states, which
are annihilated by all these generators, exist in {\it all } sectors of these
models with fixed even fermion  number.  A state in the 4-fermion sector of
the Bianchi type IX model is considered as a specific example, which
satisfies the `no-boundary' condition of Hartle and Hawking. The conclusion
is that supersymmetric minisuperspace models have a much richer manifold of
physical states than had been recognized before.
\end{abstract}

While minisuperspace models have long served as a useful laboratory for new
ideas in quantum gravity, it was only rather recently that also the quantum
theory
of {\it supersymmetric} minisuperspace models has attracted a broad interest
(see e.g. \cite{4}-\cite{16}) . It was found \cite{7} that supersymmetric
versions have special explicit analytical solutions in the {\it empty} and
{\it filled} fermion sectors which can
be interpreted as  wormhole states \cite{9} (and with a different
homogeneity condition for the gravitino, as Hartle-Hawking no-boundary states
\cite{16}). However,  even more recently it has been shown by a simple
scaling argument, that the special solutions found in the minisuperspace
models have no direct counterpart in 4-dimensional supergravity, because
there states in the empty (and also in the filled) sector cannot exist
\cite{17} . In this light another result found some time ago \cite{8} and
independently confirmed by several  authors \cite{9,10,11,12} appears very
puzzling: namely that the solutions in the empty and filled fermion sectors
are the {\it only} physical states satisfying all constraints for all
supersymmetric minisuperspace models of Bianchi type in class A \cite{19}
(without matter-coupling and with the exception,
with a certain operator ordering \cite{11,12}, of Bianchi type I). For
anisotropic supersymmetric minisuperspace models with a non-vanishing
cosmological constant
\cite{13,14,15} it even is reported in the literature that {\it no} physical
states exist, at all. Taken together these results imply that the physical
states found in the minisuperspace models have no counterpart in the full
theory, and vice-versa, which would render supersymmetric minisuperspace
models useless as models of full supergravity, contrary to the situation in
pure gravity. So, why this lecture?

Because there is a paradox: How can the supersymmetric models be more
constrained than the corresponding models in pure gravity, even though
supergravity certainly has {\it more} physical degrees of freedom than
gravity, not less (namely, in full supergravity, per space-point there are
two additional physical degrees of freedom of the
gravitino, which corresponds to two additional degrees of freedom in the
minisuperspace models).

The work \cite{25} discussed in this lecture has the purpose to clarify
these issues for the supersymmetric models of Bianchi type within class A of
the classification
of Ellis and  McCallum \cite{19}, restricting ourselves to the case of a
vanishing cosmological constant.

The starting point is the Lagrangean of $N=1$ supergravity given in
\cite{20a}. We adopt all spinor conventions given there and shall also make
free use of the excellent account of the Hamiltonian form of $N=1$
supergravity in the metric representation given in \cite{21} . In the metric
representation, which we shall use, the independent variables are taken to be
the tetrad components ${e_p}^a$ (with Einstein
indices $p=1,2,3$ from the middle of the alphabet and Lorentz indices
$a=0,1,2,3$
from the beginning of the alphabet) and the Grassmannian components
${\psi_p}^\alpha$ and their adjoints
${\bar{\psi}_p}^{\hphantom{p}\dot{\alpha}}$ (with spinor index $\alpha=1,2$;
$\dot{\alpha}=1,2$) of the Rarita-Schwinger field. The ${e_p}^a$ form the
metric tensor $h_{pq}={e_p}^ae_{qa}$ on the space-like homogeneity 3-surfaces
in the symmetric basis of 1-forms $\Bomega^p$, satisfying in Bianchi type
models of class A
\begin{equation}
\label{eq:1}
  d\Bomega^p =
\frac{1}{2}\frac{m^{pq}}{h^{1/2}}\epsilon_{qrs}\Bomega^r\wedge\Bomega^s
\end{equation}
Here $h=\det h_{pq}$ and $\epsilon_{qrs}$ denotes the components of the
Levi-Civita tensor and we shall use the notation $ V = \int
\Bomega^1\wedge\Bomega^2\wedge\Bomega^3$. The constant symmetric matrix
$m^{pq}$ is fixed by the chosen Bianchi type \cite{22}. It is invariant and
transforms as a tensor under
all coordinate changes from one {\it symmetric} basis to another one. Due
to the choice of a symmetric basis the ${e_p}^a$ and ${\psi_p}^\alpha$,
${\bar{\psi}_p}^{\dot{\alpha}}$ are functions of time only.

 From the Lagrangean one defines, as usual, the generalized
momenta $\hat{p}^p_{\hphantom{p}a}$ and $\hat{\pi}^p_{\hphantom{p}\alpha}$,
 ${\hat{\bar{\pi}}}^{\hphantom{p}p}_{\dot{\alpha}}$
of ${e_p}^a$ and ${\psi_p}^\alpha$,
${\bar{\psi}_p}^{\hphantom{\dot{\alpha}}\dot{\alpha}}$ respectively.
The Poisson brackets must be replaced by Dirac brackets due to the
appearance and subsequent elimination of second class constraints,
by which the adjoint quantities
${\bar{\psi}_p}^{\hphantom{p}\dot{\alpha}}$,
${\hat{\bar{\pi}}^p}_{\hphantom{p}\dot{\alpha}}$ are eliminated. The Dirac
brackets are decoupled by the introduction of new ${{p_+}^p}_a$ and
${\pi^p}_\alpha$.

The only non-vanishing Dirac brackets then are
\begin{eqnarray}
\{{e_p}^a, {{p_+}^q}_b\}^* &=& {\delta_p}^q{\delta_b}^a\nonumber\\
 \{{\pi_p}^\alpha, {\psi^q}_\beta\}^* &=&
   -{\delta_p}^q{\delta_\beta}^\alpha\,.
\end{eqnarray}
The supersymmetry generators $S_\alpha$, $\bar{S}_{\dot{\alpha}}$ and Lorentz
generators $J_{\alpha\beta}$, $\bar{J}_{\dot{\alpha}\dot{\beta}}$ in this
representation are obtained as
\begin{eqnarray}
\label{eq:6}
  S_\alpha &=& - {\cal C}_{pr}^{\dot{\alpha}\beta}
            \left(\frac{1}{2}Vm^{pq}{e_q}^a+{\textstyle\frac{i}{2}}
                 {p_+}^{pa}\right)
   \sigma_{a\alpha\dot{\alpha}}
  {\pi^r}_\beta\nonumber\\
  \bar{S}_{\dot{\alpha}} &=& \left(\frac{1}{2}Vm^{pq}{e_q}^a
     -{\textstyle\frac{i}{2}}{p_+}^{pa}\right)
    \sigma_{a\alpha\dot{\alpha}}{\psi_p}^\alpha
\end{eqnarray}
and
\begin{eqnarray}
\label{eq:7}
  J_{\alpha\beta} &=& +{\textstyle\frac{1}{2}}
       (\sigma^{ac}\epsilon)_{\alpha\beta}
    \left(e_{pa}{{p_+}^p}_c-e_{pc}{{p_+}^p}_a\right)
  \nonumber\\
      && -{\textstyle\frac{1}{2}}
         \left(\psi_{p\alpha}{\pi^p}_\beta+\psi_{p\beta}
           {\pi^p}_\alpha\right)
  \nonumber\\
  \bar{J}_{\dot{\alpha}\dot{\beta}} &=& -{\textstyle\frac{1}{2}}
    (\epsilon\bar{\sigma}^{ac})_{\dot{\alpha}\dot{\beta}}
     \left(e_{pa}{{p_+}^p}_c-e_{pc}{{p_+}^p}_a\right)\,.
\end{eqnarray}
with
\begin{equation}
\label{eq:4}
  {\cal C}_{pq}^{\dot{\alpha\alpha}} = -{\textstyle\frac{1}{2Vh^{1/2}}}
    \left(ih_{pq}n^a-\varepsilon_{pqr}e^{ra}\right)
       \bar{\sigma}_a^{\hphantom{a}\dot{\alpha}\alpha}\,.
\end{equation}

Canonical quantization is achieved by putting
\begin{equation}
    {{p_+}^p}_a=-i\hbar(\partial /\partial{e_p}^a) \,\qquad
   {\pi^p}_\alpha = -i\hbar(\partial /\partial{\psi_p}^\alpha)\,.
\end{equation}
In $S_\alpha$ there is then an ordering ambiguity between ${{p_+}^p}_a$ and
${\cal C}_{pr}^{\dot{\alpha}\beta}$, which is resolved by adopting
the ordering as written in
eq.~(\ref{eq:6}) \cite{20}. Now a somewhat lengthy calculation gives the
following generator algebra \cite{25,20}: The commutators of
$J_{\alpha\beta}$, $\bar{J}_{\dot{\alpha}\dot{\beta}}$ with any operator is
determined by its changes under infinitesimal Lorentz transformations.
The remaining commutators and anti-commutators are
\begin{eqnarray}
\label{eq:9}
\Big[S_\alpha,S_\beta\Big]_+&=&0=
 \Big[\bar{S}_{\dot{\alpha}},\bar{S}_{\dot{\beta}}\Big]_+\\
\label{eq:10}
  \Big[S_{\alpha},\bar{S}_{\dot{\alpha}}\Big]_+&=&
    - \frac{\hbar}{2}H_{\alpha\dot{\alpha}}\\
\label{eq:16}
  \Big[H_{\alpha\dot{\alpha}},S_{\beta}\Big]_- &=&
    -i\hbar\varepsilon_{\alpha\beta}
       {\bar{D}_{\dot{\alpha}}}^{\hphantom{p}{\dot{\beta}\dot{\gamma}}}
        \bar{J}_{\dot{\beta}\dot{\gamma}}\\
\label{eq:17}
   \Big[H_{\alpha\dot{\alpha}},\bar{S}_{\dot{\beta}}\Big]_- &=&
      i\hbar\varepsilon_{\dot{\alpha}\dot{\beta}}{J}_{\beta\gamma}
 {D_\alpha}^{\beta\gamma}\nonumber\\
    &=&
i\hbar\varepsilon_{\dot{\alpha}\dot{\beta}}\Big[{D_\alpha}^{\beta\gamma}
{J}_{\beta\gamma}+i\hbar{\bar{E}_\alpha}^{\phantom{p}\dot{\gamma}
\dot{\delta}}
\bar{J}_{\dot{\gamma}\dot{\delta}}+\frac{i\hbar n^a}{Vh^{1/2}}
   \sigma_{a\alpha\dot{\gamma}}\bar{S}^{\dot{\gamma}}
\Big]\,.
\end{eqnarray}
The operator $H_{\alpha\dot{\alpha}}$ is here {\it defined} by the
anticommutator (\ref{eq:10}). We checked that it differs from the
diffeomorphism --- and Hamiltonian constraints
$\tilde{H}_{\alpha\dot{\alpha}}
=\sigma_{a\alpha\dot{\alpha}}({e_p}^aH^p+n^aH) $ only by terms proportional
to Lorentz generators. The commutators
$[\bar{S}_{\dot{\beta}}$, $H_{\alpha\dot{\alpha}}]_-$ and
$[S_{\beta}, H_{\alpha\dot{\alpha}}]_-$ are the essential new results which
will turn out to be crucial, in the following. ${D_\alpha}^{\beta\gamma}$,
${\bar{E}_\alpha}^{\hphantom{p}\dot{\gamma}\dot{\delta}}$ are Grassmann-odd
operators whose explicit form we shall not need. Eqs.
(\ref{eq:9})-(\ref{eq:17}) demonstrate explicitely that the algebra of the
generators closes, and thereby confirm the assumption made in earlier work on
the same minisuperspace models.

The physical states of these models are given by all the states which are
annihilated by the generators $S_\alpha$,
$\bar{S}_{\dot{\alpha}}$,
$J_{\alpha\beta}$, $\bar{J}_{\dot{\alpha}\dot{\beta}}$,
$H_{\alpha\dot{\alpha}}$.
The Lorentz generators automatically annihilate all states which are
Lorentz scalars. Therefore, it is sufficient to demand that physical
states are Lorentz scalars and annihilated by $S_\alpha$ and
$\bar{S}_{\dot{\alpha}}$; their annihilation by $H_{\alpha\dot{\alpha}}$
is then automatically guaranteed by the generator algebra. The form of the
constraint operators guarantees that physical states have a fixed fermion
number $F={\psi_p}^\alpha\partial/\partial\psi^{\alpha}_p$,
which must be an even number in Lorentz-invariant
states and ranges from 0 to 6 in the present models.

The physical states in the sectors $F=0$ and $F=6$ are easily
obtained, and are, respectively, given by \cite{20}
\begin{eqnarray}
\label{eq:19}
  \Psi_0 &=& {\mbox{\rm const}\,} e^{\frac{V}{2\hbar}m^{pq}h_{pq}}
             \nonumber\\
  \Psi_6 &=& {\mbox{\rm const}\,} e^{-\frac{V}{2\hbar}m^{pq}h_{pq}}
     \prod_r\left(\psi_r\right)^2
\end{eqnarray}
reproducing a well known result \cite{7}-\cite{12}.

In order to show that there exist physical states in the 2-fermion sector
as well we consider the wave-function \cite{25}
\begin{equation}
\label{eq:20}
    \Psi_2 = \bar{S}_{\dot{\alpha}}\bar{S}^{\dot{\alpha}}f(h_{pq}),
\end{equation}
with $\bar{S}^{\dot{\alpha}}f\neq0$, where $f$ is a function of the $h_{pq}$
only, and therefore, like
$\bar{S}_{\dot{\alpha}}\bar{S}^{\dot{\alpha}}$, a Lorentz scalar.
Therefore
$\Psi_2 $ {\it automatically} satisfies the Lorentz
constraints and the $\bar{S}$-constraints
because of eq.~(\ref{eq:9}). The only remaining condition is
$S_\alpha\Psi_2 = 0$, which, after the use of eqs.~(\ref{eq:20}),
(\ref{eq:10}), reduces to
\begin{equation}
\label{eq:21}
  \Big[H_{\alpha\dot{\alpha}},\bar{S}^{\dot{\alpha}}\Big]_-f \quad
   +2\bar{S}^{\dot{\alpha}}H_{\alpha\dot{\alpha}}f \quad =0
\end{equation}
The first term, thanks to eq. (\ref{eq:17}), reduces to terms proportional
to $J_{\beta\gamma}$, $\bar{J} _{\dot{\gamma}\dot{\delta}}$,
$\bar{S}^{\dot{\gamma}}$.
The terms with $J_{\beta\gamma}$, $\bar{J} _{\dot{\gamma}\dot{\delta}}$
vanish because $f$ is a Lorentz scalar.
In the remainder $\bar{S}^{\dot{\gamma}}$ can be factored out to the left,
using
$[\bar{S}^{\dot{\gamma}},\sigma _{a\alpha\dot{\gamma}} n^a/h^{1/2}]_- = 0$.
Then it is seen that eq. (\ref{eq:21}) is solved if $f$ satisfies the
Wheeler-DeWitt equation \cite{20}
\begin{equation}
\label{eq:23}
 \left({H_{\alpha\dot{\alpha}}}^{(0)}-\frac{\hbar^2}{Vh^{1/2}}n^a
\sigma _{a\alpha\dot{\alpha}}\right)f(h_{pq})=0
\end{equation}
where ${H_{\alpha\dot{\alpha}}}^{(0)}$ consists only of the bosonic terms
of $H_{\alpha\dot{\alpha}}$, i.e. of the terms which remain if
${\pi^p}_\alpha$
is brought to the right and then equated to zero.
Any solution of this
Wheeler-DeWitt equation, which may be specified further by imposing boundary
conditions, determines a solution in the 2-fermion sector via
eq.~(\ref{eq:20}), with
a definite dependence on the fermionic variables. It should be noted that
the norm of $\Psi_2$ vanishes due to the appearance of
$\bar{S}_{\dot{\alpha}}$
in eq.~(\ref{eq:20}) and the fact that $S_\alpha$
is the adjoint of $\bar{S}_{\dot{\alpha}}$.
However, including the necessary gauge-fixing condition in the measure of
the scalar product, the norm of $\Psi_2$ (and of $\Psi_4$ to be considered
below)
will not vanish.

We now
turn to physical states in the 4-fermion sector. Similarly to
eq.~(\ref{eq:20}) the wave-function
\begin{equation}
\label{eq:24}
  \Psi_4 = S^\alpha S_\alpha g(h_{pq})\prod_{r=1}^3(\psi_r)^2\,.
\end{equation}
with $\Psi_4\neq0$ automatically satisfies the Lorentz constraints and the
$S_\alpha$-constraint.
It remains to satisfy the $\bar{S}_{\dot{\alpha}}$ constraint.  A
calculation  similar to the one leading from eq. (\ref{eq:20}) to eq.
(\ref{eq:24}) then gives \cite{25,20} the
Wheeler-DeWitt equation.
\begin{equation}
\label{eq:26}
  {H_{\alpha\dot{\alpha}}}^{(1)}g(h_{pq}) = 0
\end{equation}
where ${H_{\alpha\dot{\alpha}}}^{(1)}$ consists of those terms of
$H_{\alpha\dot{\alpha}}$ which remain if the ${\pi^p}_a$ are brought
to the left and then equated to zero.  Eq.  (\ref{eq:26}) is
slightly different from that obeyed by the amplitude in the 2-fermion
sector, but the degree of generality of the solution is
the same.

Our results differ from earlier work on the supersymmetric Bianchi models
in class A \cite{8}-\cite{12} which concluded that physical states in
the 2- and 4-fermion sectors do not exist.  In fact the ansatz made there
was overly restrictive and did not include all the Lorentz invariants
contained in the states
(\ref{eq:20}) and (\ref{eq:24}).

What about the relation to full supergravity?  Provided the algebra
of the local generators of the constraints still has a form like
eqs.~(\ref{eq:9})-(\ref{eq:17}) physical states still exist which look
somewhat like $\Psi_2$ and $\Psi_4$, but contain formal products of
$(\bar{S})^2$ or $(S)^2$ over all points of the space-like 3-surface, thus
leading to states with infinite fermion number. Such states have recently
been discussed in \cite{17}. The new physical states discussed in the
present
paper are the direct minisuperspace analogues of such states in full
supergravity; even though, due to the reduction to minisuperspace the
fermion number is, of course, finite. The new states display the same
richness of gravitational
dynamical behavior as the Bianchi models in pure gravity. In this respect
they differ qualitatively from the earlier found states in the empty and
the
full fermion sectors, which are highly symmetric and do not describe the
very rich  dynamical behavior of the Bianchi models in the classical limit.
The new physical states described here
span an infinite-dimensional Hilbert space just as the states of the Bianchi
models of
pure gravity. Just how this Hilbert space ought to be constructed remains
one of the interesting open questions.

Before ending let us consider a specific example, namely the case of Bianchi
type IX. There the metric $h_{pq}$ can be parametrized by the Euler angles
$\varphi,\vartheta,\chi$ contained  in a rotation matrix $\Omega_{pi}$ and
three scale parameters
\begin{equation}
\label{eq:24a}
  \left( e^{2\beta}\right)_{ij}=
e^{2\alpha}\diag\left(e^{2\beta_++2\sqrt{3}\beta_-},
    e^{2\beta_+-2\sqrt{3}\beta_-},
     e^{-4\beta_+}\right)
\end{equation}
via $h_{pq}=\Omega_{pi}(e^{2\beta})_{ij}\Omega_{qj}$.

Let us consider, in particular, eq.(\ref{eq:26}) for the 4-fermion sector.
The three equations \newline${e_p}^a\bar{\sigma}_a^{\dot{\alpha}\alpha}
{H_{\alpha\dot{\alpha}}}^{(1)}g=0$ turn out to be automatically satisfied if
$g$ is {\it independent}
of the Euler angles $\varphi,\vartheta,\chi$. The only remaining constraint
$n^a\bar{\sigma}_a^{\dot{\alpha}\alpha}{H_{\alpha\dot{\alpha}}}^{(1)}g=0$
then takes the form
\begin{eqnarray}
\label{eq:25a}
&&  \Bigg[
     -\frac{\hbar^2}{V^2}\left(\frac{\partial}{\partial\alpha}\right)^2+
\frac{\hbar^2}{V^2}
       \left(\frac{\partial}{\partial\beta_+}\right)^2+\frac{\hbar^2}{V^2}
        \left(\frac{\partial}{\partial\beta_-}\right)^2+
         \left(\frac{\partial\phi}{\partial\alpha}\right)^2-
          \left(\frac{\partial\phi}{\partial\beta_+}\right)^2-
           \left(\frac{\partial\phi}{\partial\beta_-}\right)^2\nonumber\\
&& \mbox{\hspace{2.5cm}}
         +\frac{\hbar}{V}\left(-\frac{\partial^2\phi}{\partial\alpha^2}+
             \frac{\partial^2\phi}{\partial\beta_+^2}+
              \frac{\partial^2\phi}{\partial\beta_-^2} \right)\Bigg]
               g(\alpha,\beta_+,\beta_-)=0
\end{eqnarray}
with
\[
 \phi=\frac{1}{2}e^{2\alpha}\left(2e^{2\beta_+}\cosh2\sqrt{3}
\beta_-+e^{-4\beta_+}\right)\,.
\]
This equation is of exactly the same form as given in ref. \cite{7}. Even
though $g=\exp(-V\phi/\hbar)$ solves eq.~(\ref{eq:25a}) exactly \cite{7},
this solution is not permitted here, because it would fail to satisfy the
condition $\Psi_4\neq 0$. Remarkably, however, {\it another} equally simple
exact solution of eq.~(\ref{eq:25a}) exists without this defect. It is
\begin{eqnarray}
&& g=\const\exp\Big[-\frac{V}{2\hbar}e^{2\alpha}\big(2e^{2\beta_+}
     (\cosh 2\sqrt{3}\beta_--1)\nonumber\\&&
\mbox{\hspace{4.3cm}}+e^{-4\beta_+}-4e^{-\beta_+}
  \cosh\sqrt{3}\beta_-\big)\Big]\,.
\end{eqnarray}
As has been discussed elsewhere (see e.g. the last reference \cite{7}), this
state satisfies the requirements of a Hartle-Hawking `no-boundary' state. In
Bianchi type IX models with the homogeneity condition
$\Bpsi^\alpha={\psi_p}^\alpha(t)\Bomega^p$ such a state, up til now, was
mysteriously absent \cite{9} which led to the proposal to modify the
homogeneity condition for the Rarita-Schwinger field \cite{16}. It is
therefore very interesting and gratifying to see now such a state appear
among the new states (\ref{eq:24}) without any modification in the
homogeneity condition.

In summary this work demonstrates that anisotropic supersymmetric
minisuperspace models have a much richer manifold of physical states than had
been recognized up to now with clear correspondences with states both in
minisuperspace models of pure gravity and in full supergravity.

This work has been supported by the Deutsche Forschungsgemeinschaft through
the Sonderforschungsbereich 237 ``Unordnung und gro{\ss}e Fluktuationen''.
One of us (A.~Csord\'as) would like to acknowledge additional support by
The Hungarian National Scientific Research Foundation under Grant number
F4472.

\end{document}